\documentclass[twocolumn,prl]{revtex4-1}
\usepackage{epsfig,amssymb,amsmath,epstopdf}

\usepackage[english]{babel}

\begin{document}
\title{Re-orientation of easy axis in $\varphi_0$ junction}

\author{Yu. M. Shukrinov$^{1,2}$, A. Mazanik$^{1,3}$, I. R. Rahmonov$^{1,4}$, A. E. Botha$^{5}$, A. Buzdin$^{6}$}

\affiliation{$^{1}$	BLTP, JINR, Dubna, Moscow Region, 141980, Russia\\
	$^{2}$Dubna State University, Dubna, Russian Federation\\
$^{3}$Moscow Institute of Physics and Technology, Institutsky lane 9, Dolgoprudny, Moscow region, 141700, Russian Federation\\
  $^{4}$Umarov Physical Technical Institute, TAS, Dushanbe, 734063, Tajikistan\\
 $^5$ Department of Physics, University of South Africa, Private Bag X6, Florida 1710, South Africa\\
 $^{6}$ University Bordeaux, LOMA UMR-CNRS 5798, F-33405 Talence Cedex, France}
\date{\today}
\date{\today}

\begin{abstract}
We study theoretically a dynamics of $\varphi_0$ junction with direct coupling   between magnetic moment and Josephson current which shows features close to  Kapitza pendulum. We have found that starting with oscillations along $z$-axis, the character of magnetization  dynamics  changes crucially  and stable position of magnetic moment $\vec m$ is realized between $z-$ and $y$-axes depending on parameters of the system. Changes in critical current and spin-orbit interaction lead to the different stability regions for magnetization. An excellent agreement between analytical and numerical results is obtained for low values of the Josephson to magnetic energy ratio.
\end{abstract}
\keywords{Superconducting electronics, $\varphi_0$-junction, magnetization reversal, spin-orbit interaction}

\maketitle

%\begin{widetext}

%\subsection{Introduction}
%{\it Introduction --}
A ferromagnet makes a strong influence on Josephson junction (JJ), particularly, superconductor-ferromagnet-superconductor (SFS) JJ oscillates between $0-$ and $\pi-$junctions  as thickness of ferromagnet increases \cite{ryazanov01, ryazanov06,robinson05}. In a closed superconducting loop with $\pi$-junction a spontaneous circulating current appears without applied magnetic flux \cite{bulaevsky77}. Very special situation is possible when the weak link is a non-centrosymmetric magnetic metal with broken inversion symmetry like $MnSi$ or $FeGe$. In this case, Rashba type spin-orbit coupling leads to a general current-phase relation $j=j_c\sin(\varphi-\varphi_0)$, where $\varphi_0$ is proportional to the strength of the broken inversion symmetry magnetic interaction \cite{buzdin08,krive,reynoso}. An important issue is that the  $\varphi_0$-junction provide a mechanism of a direct coupling between suppercurrent (superconducting phase) and magnetic moment.

The mechanical analog with a simple pendulum with applied torque occurred to be very useful to get a better insight into a physics of Josephson devices \cite{barone}. Here we introduce an another analog between magnetic $\varphi_0$-junction and a pendulum with oscillating point of suspension. A particle moving simultaneously in the permanent field and in the field oscillating with a high frequency, demonstrates a very interesting feature: new stability points appear at some parameters of the particle and fields \cite{kapitza,landau}. The system, where such feature is realized, usually is called as Kapitza pendulum. Particularly, in pendulum with vibrating point of suspension (pivot point), the external sinusoidal force can invert the stability position of the pendulum \cite{kapitza}. In this case the bottom equilibrium position is no longer stable. Any tiny deviation from the vertical increases in amplitude with time. In the presence of a periodic drive, the unstable fixed point can become dynamically stable. Kapitza provided an analytical insight into the reasons of stability by splitting the motion into ``fast'' and ``slow'' variables and by introducing an effective potential.  This result was first obtained by Kapitza in the high frequency limit. By averaging the classical equations of motion over the fast oscillations of the drive, Kapitza found that the ``upper'' extremum becomes stable for large enough driving amplitudes. This pioneering work initiated the field of vibrational mechanics, and the Kapitza method is used for description of periodic processes in different physical systems such as atomic physics, plasma physics, and cybernetical physics (see \cite{citro2015,boukobza2010} and references therein). In nonlinear control theory the Kapitza pendulum is used as an example of a parametric oscillator that demonstrates the concept of ``dynamic stabilization''.

In this paper we study dynamics of $\varphi_0$ junction with direct coupling   between magnetic moment and Josephson current and found that applying the external voltage may lead to the different stability regions for magnetization. We have found the manifestation of Kapitza pendulum features in the $\varphi_0$-junction \cite{buzdin05,konschelle09}. Developing our previous approach \cite{apl2017}, we have investigated the effect of superconducting current on the dynamics of magnetic momentum.  We show that starting with oscillations along $z$-axis, the character of $\vec m$ dynamics  changes crucially  and stable position of $\vec m$ becomes between $z-$ and $y$-axes depending on parameters of the system.

The ac Josephson effect provides an ideal tool to study magnetic dynamics
in a $\varphi_{0}$--junction. To realize the ac Josephson effect, we apply the constant voltage $V$ to the considered
$\varphi_{0}$--junction. In such a case, the superconducting phase varies
with time like $\varphi(t) = \omega_{J} t$, where
$\omega_{J}=2eV/\hbar$ is Josephson frequency \cite{konschelle09,josephson}. Dynamics of considered system is described by the Landau-Lifshitz-Gilbert equation
\begin{equation}
\frac{d {\bf M}}{dt}=\gamma {\bf H_{eff}} \times {\bf M}+\frac{\alpha}{M_{0}}\bigg({\bf M}\times \frac{d {\bf M}}{dt} \bigg)
\label{momentum}
\end{equation}
with effective magnetic field ${\bf H_{eff}}$ in the form \cite{konschelle09}
\begin{equation}
{\bf H_{eff}}=\frac{K}{M_{0}}\bigg[\Gamma \sin\bigg(\omega t - \varphi_0 \bigg) {\bf\widehat{y}} + \frac{M_{z}}{M_{0}}{\bf\widehat{z}}\bigg]
\label{effective_field}
\end{equation}
where $\gamma$ -- gyromagnetic ratio, $\alpha$ -- phenomenological damping constant, $M_{0}=\|{\bf M}\|$, $M_{i}$ are components of ${\bf M}$, and we will use normalized units below $m_i=\frac{M_i}{M_0}$.  Here $\varphi_0=r \frac{M_{y}}{M_{0}}$, $r=l\upsilon_{so}/\upsilon_{F}$, $l=4 h L/\hbar \upsilon_{F}$, $L$--length of $F$ layer, $h$--exchange field of the $F$ layer, $\displaystyle \Gamma=Gr$, $\displaystyle G=E_{J}/(K \mathcal{V})$, $\displaystyle E_{J}=\Phi_{0}I_{c}/2\pi$ is the Josephson energy. Here $\Phi_{0}$
is the flux quantum, $I_{c}$ is the critical current, $\upsilon_{F}$ is Fermi velocity, the
parameter $\upsilon_{so}/\upsilon_{F}$ characterizes a relative
strength of spin-orbit interaction, $K$ is the anisotropic constant,
and $\mathcal{V}$ is the volume of the $F$ layer. To investigate the dynamics of considered system numerically, we write the equation (\ref{momentum}) in the dimensionless form (see system of equations (1) in the Supplement ). In that system time is normalized to the inverse ferromagnetic resonance frequency $\omega_{F}=\gamma K/M_{0}: (t\rightarrow t \omega_F)$, so $\omega$ is normalized to the $\omega_F$.

\begin{figure}[h!]
 \centering
  \includegraphics[height=38mm]{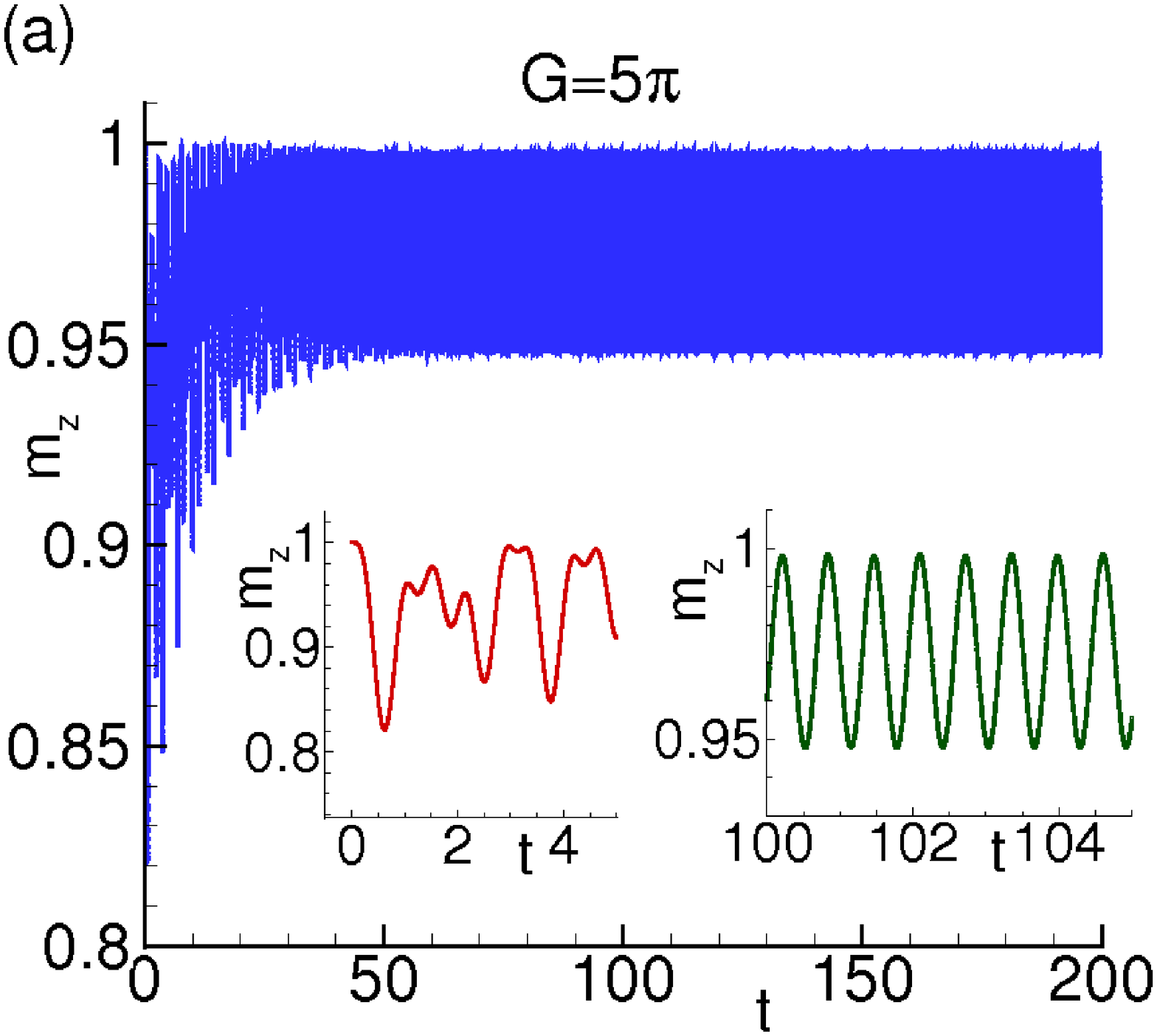} \includegraphics[height=38mm]{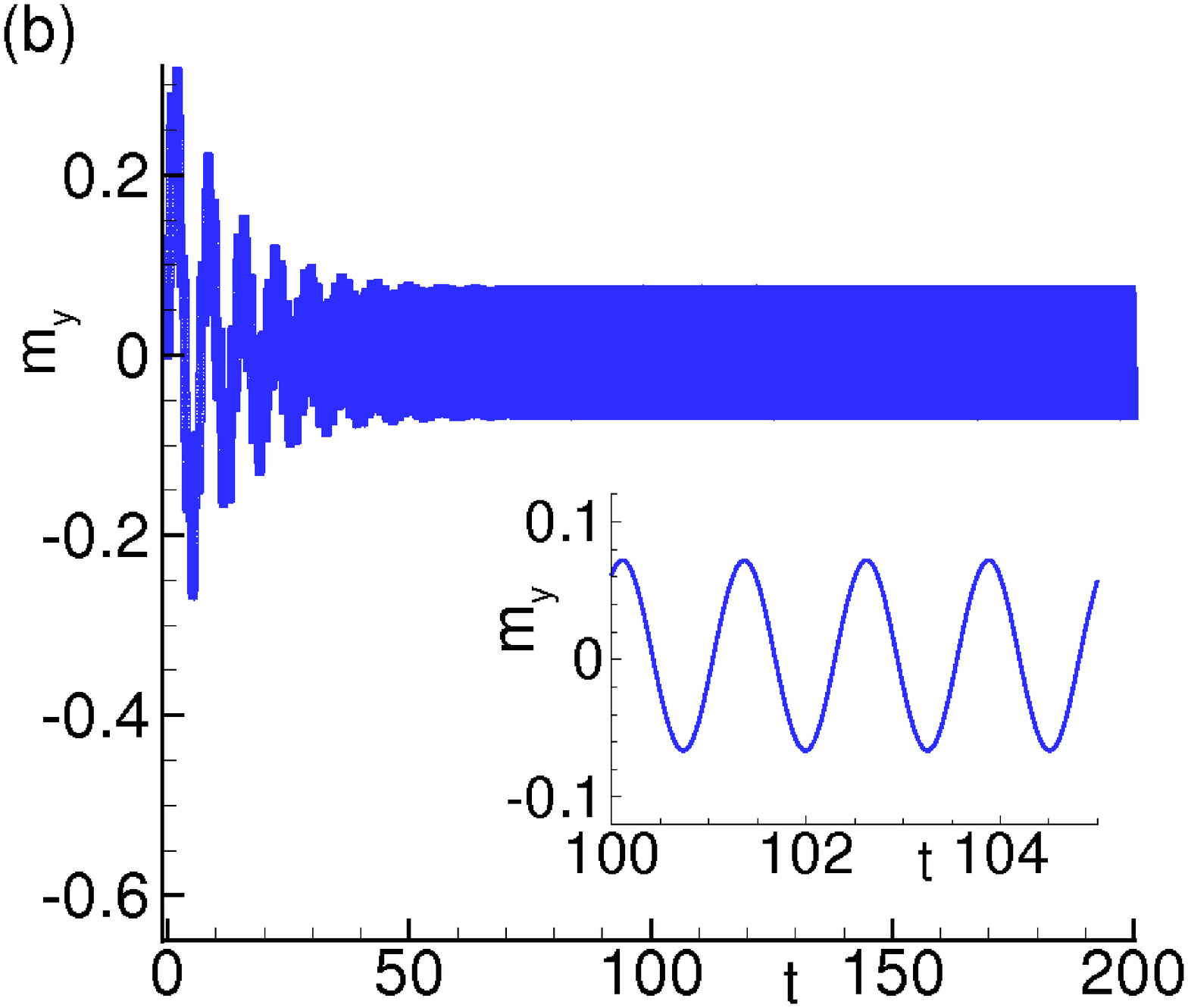}
 \includegraphics[height=38mm]{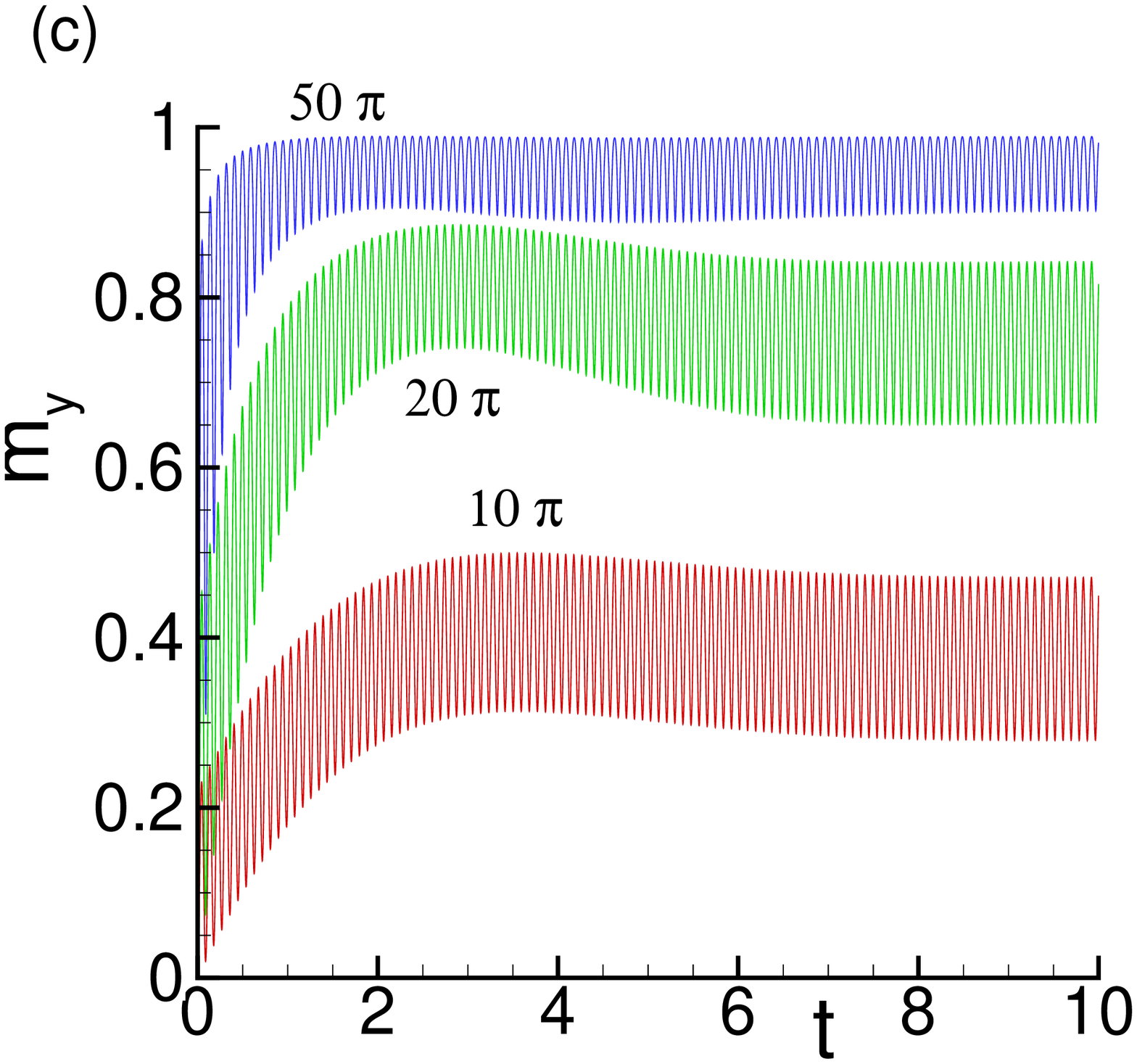} \includegraphics[height=38mm]{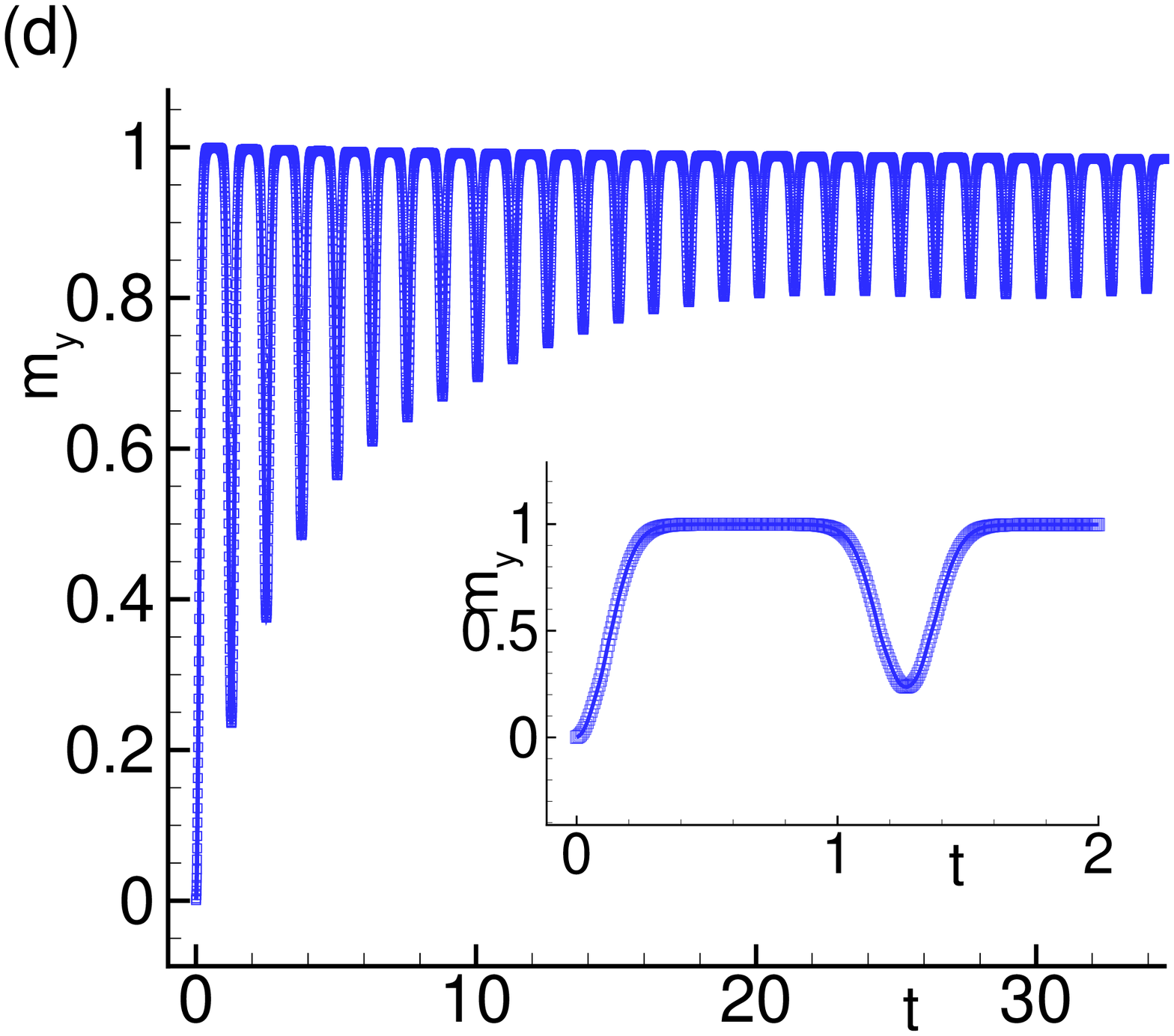}
\caption{(a) Dynamics of $m_z$ component at $G=5\pi$, $r=0.1$; (b) The same for $m_y$; (c) Effect of G at $r=0.5$. The numbers indicate the $n$ value in $G=n\pi$. (d) Dynamics of $m_y$ component at $G=400\pi$, $r=0.1$.  In all figures $\alpha=0.1$, $\omega=5$.}
\label{1}
\end{figure}

\begin{figure}[h!]
\centering
\includegraphics[scale=0.45]{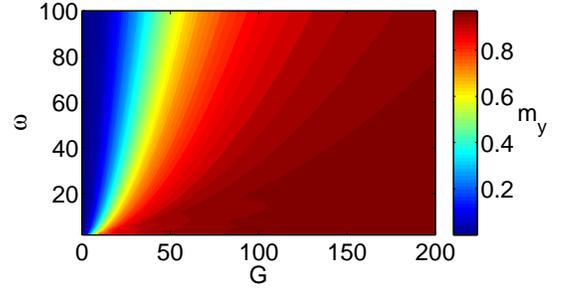}
\caption{The $\omega-G$-diagram for averaged $m_y$ shown by color.}
    \label{2}
\end{figure}

First we present results of numerical simulations of system according to equation (\ref{momentum}). Figure \ref{1} shows dynamics of magnetization components $m_{z}$
and $m_{y}$ at different parameter $G$ demonstrating the re-orientation of the oscillations around $z$ axis to the oscillations around $y$ axis.
%We see that $m_{z}$ jumps from value $m_{z}=1$ and oscillates around $m_z=0$ demonstrating periodically some splashing.
With increase in $G$, the component $m_y$ goes from zeroth to $m_y=1$.
Figure \ref{1}(a) demonstrates the time dependence of $m_z$ at $G=5\pi$. The character of oscillations in the beginning and in the middle of time interval is shown in the insets. We see that the average value of $m_{z}$ deviates from one. Figure \ref{1}(b) shows the corresponding oscillation of $m_{y}$. Figure \ref{1}(c) demonstrates a stabilization of $m_y$ oscillations around of some average values of $m_y$  between $z$- and $y$- directions with increase in $G$, shows the re-orientation of oscillations at three values of $G=10\pi, 20\pi, 50\pi$. With an increase in $G$,  time of re-orientation from z-direction to y-direction is decreased essentially. At enough large value of $G$ (see $G=400\pi$ in Fig.1(d)), the average value of $m_{y}$ is getting close to $1$. The oscillations show periodically splashing related to the Josephson frequency. Inset demonstrates the jumps of $m_y$ from the value $m_y=1$ with Josephson frequency. The amplitude of jumps decreases in time and it is getting smaller and in a shorter time with an increase in $G$. We note that the effect of $r$ leads to the similar features as the changes in $G$.
So, the situation is reminiscent of Kapitza pendulum (a pendulum whose point of suspension vibrates) where the external sinusoidal force can invert the stability position of the pendulum~\cite{kapitza}.
Variation of average $m_{y}$ as a function of frequency $\omega$ and $G$ is shown in Fig.\ref{2}. We see that an increase in $G$ makes orientation of $m_y$ along y-axis stable, but frequency dependence differs from characteristic Kapitza pendulumn behavior, as we can see below from the results presented in Fig.\ref{4}.
The system of equations (\ref{momentum}) describing dynamics of magnetic moment  in angular variables $m_z =\cos{\theta}$, $ m_x = \sin{\theta}\cos{\phi}$, $m_y = \sin{\theta}\sin{\phi}$ can be written as
\begin{equation}\label{eq:main}
    \begin{aligned}
        \begin{cases}
            \dot{\phi} = \frac{\cos{\theta}}{1+\alpha^2} - \\
            \frac{Gr}{1 + \alpha^2}\frac{1}{\sin{\theta}}\left[ \cos{\theta} \sin{\phi} - \alpha \cos{\phi}\right]\sin{\left[ \omega\tau - r \sin{\theta} \sin{\phi} \right]},\\
            \dot{\theta} =- \frac{\alpha \sin{2\theta}}{2(1+\alpha^2) } + \\
            \frac{Gr}{1+\alpha^2}\left[ \alpha \cos{\theta} \sin{\phi} +  \cos{\phi} \right] \sin{\left[\omega \tau - r \sin{\theta}\sin{\phi}\right]}.
        \end{cases}
    \end{aligned}
\end{equation}
In the case of absence of coupling between the magnetic moment and the Josephson junction at $G = 0$
\begin{equation} \label{eq:fr}
    \begin{cases}
        \dot{\phi} = \frac{\cos{\theta}}{1 + \alpha^2}, \\
        \dot{\theta} = -\frac{\alpha\sin{2\theta}}{2(1+\alpha^2)}.
    \end{cases}
\end{equation}
the solution has a form
\
\begin{equation}
     \theta(\tau) = \operatorname{ArcTan}{\left[\exp{\left\{ \log{\tan{\theta_0}} - \frac{\alpha \tau}{1 + \alpha^2}  \right\}}\right]},
\end{equation}
where $\tan{\theta_0}$ is considered to be positive for simplicity.
So, the characteristic scale in time to order magnetic moment along the easy axes is  $\tau^{\star} = \frac{1 + \alpha^2}{\alpha}$. To realize the Kapitza pendulum we should investigate the case  $\omega \gg \frac{2\pi}{\tau^{\star}} = \frac{2\pi \alpha}{1+\alpha^2}$, where $\omega$ is a frequency of the external fast-varying  field.

Let us consider that
\begin{equation} \label{eq:Method}
    \begin{aligned}
        \theta \to \Theta + \xi, \\
        \phi \to \Phi + \eta.
    \end{aligned}
\end{equation}
Here $\Theta$ and $\Phi$ describe slow movement, while $\xi$ and  $\eta$ are coordinates for fast-varying movement. Conditions for  $\xi$ and $\eta$ are discussed in the Supplement. We consider $\overline{\theta} = \Theta$, $\overline{\phi} = \Phi$, where averaging is taken over the period of the fast-varying force $T = \frac{2\pi}{\omega}$.

The system of equations for slow movement has a form (see Supplement)
\begin{equation} \label{eq:main1}
    \begin{aligned}
        \begin{cases}
            \dot{\Phi} = \frac{\cos{\Theta}}{1+\alpha^2}-\frac{\left(Gr\right)^2 r \alpha}{2 \omega (1+\alpha^2)^2}\cdot\frac{1}{\sin{\Theta}}[\cos{\Theta}\sin{\Phi}\\
\hspace{1 cm}             - \alpha\cos{\Phi} ]\left\{1-\sin^2{\Theta}\sin^2{\Phi}\right\}, \\
            \dot{\Theta} =- \frac{\alpha \sin{2 \Theta}}{2(1+\alpha^2)}+\frac{\left(Gr\right)^2 r \alpha}{2 \omega (1+\alpha^2)^2}[\alpha\cos{\Theta}\sin{\Phi}\\
\hspace{1 cm}            + \cos{\Phi} ]\left\{1-\sin^2{\Theta}\sin^2{\Phi}\right\}
        \end{cases}
    \end{aligned}
\end{equation}
There are some equilibrium points of the system following from $\dot{\Theta} = 0, \dot{\Phi} = 0$.
\begin{figure}[h!]
 \centering
\includegraphics[height=50mm]{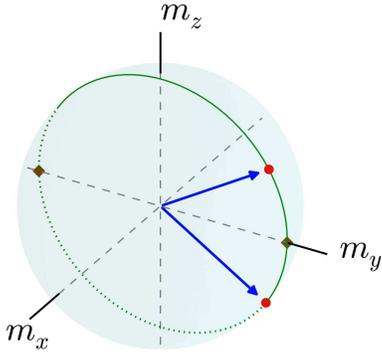}
\caption{Demonstration of equilibrium points of equation (\ref{eq:main1}). Arrows show stable points, other two are unstable.}
\label{3}
\end{figure}

The first pair is located on the equator
\begin{equation}\label{eq:ep0}
 \Theta_0 = \pi/2, \Phi_0 = \pi/2 \text{ and } \Theta_0 = \pi/2, \Phi_0 = 3\pi/2.
\end{equation}
Linearization of  (\ref{eq:main1}) after substitution $\Phi = \Phi_0 + \delta \phi,\Theta = \Theta_0 + \delta \theta$ gives
\begin{equation}
    \begin{aligned}
        \begin{cases}
            \dot{\delta \phi} =  \frac{1}{1+\alpha^2} \delta \theta, \\
            \dot{\delta \theta} = \frac{\alpha}{1+\alpha^2}\delta\theta.
        \end{cases}
    \end{aligned}
\end{equation}
It's clear that $\delta\theta \sim e^{\frac{\alpha \tau}{1 + \alpha^2}}$, so these points are unstable.

Using (\ref{eq:main1}) we find that the equilibrium points  at $\Phi_0 = \frac{\pi}{2}$ are described by equation
\begin{equation}\label{eq:ep1}
    \sin{\Theta_0} ={\frac{-1+\sqrt{1+4 \beta^2}}{2 \beta}},
\end{equation}
where $\beta = \frac{\left(Gr\right)^2 r \alpha}{2 \omega (1 + \alpha^2)}$, which can be approximated as $ \sin{\Theta_0} =\beta$ at small $\beta$. It has two solutions in the interval $0 \leq \Theta_0 \leq \pi$. Note that at $\Phi_0 = \frac{3\pi}{2}$ we have  $\sin{\Theta_0} ={\frac{1-\sqrt{1+4 \beta^2}}{2 \beta}}$  which leads to the negative $ \sin{\Theta_0}$, but $\Theta$ is always positive. So, there are no any stable points at  $\Phi_0 = \frac{3\pi}{2}$.

To find out if these points are stable we have to make linearization of equation (\ref{eq:main1}) near these points and test eigenvalues of linearized system. The straightforward calculation shows, that the real parts of eigenvalues of the corresponding system are always negative. It means that the second pair of points are always stable (see Supplement).

Results of numerical calculations of the averaged $m_y$ as a function $G$ at different frequencies are presented in Fig. \ref{4}. We see that system reminiscent the Kapitza pendulum behavior: averaged  $m_y$ component  characterizing changes of stability direction is growing  with $G$. Character of behavior depends essentially on the frequency of the fast-varying field getting very sharp at small $\omega$. Figure \ref{4} also compares the $m_y(G)$ dependence obtained by analytical and numerical calculations.  We see results at three frequencies: $\omega=0.5$, $20$  and $70$.  Analytical dependence calculated according to  the formula (\ref{eq:ep1}). Two methods are used for numerics: numerics-1 is based on standard program ``Mathematica'' (shown by triangles, calculated for $\omega=70$) and numerics-2 presents results of direct solution of system (\ref{momentum})  based on the Runge-Kutta fourth order. Both numerics gives practically the same results. An excellent agreement between analytical and numerical results is obtained at low $G$, depending on frequency of the fast-varying field. For high frequencies the coincidence is still good at rather large G.
\begin{figure}[h!]
 \centering
\includegraphics[height=60mm]{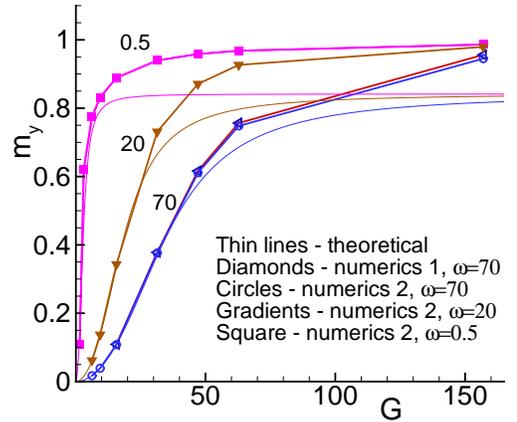}
\caption{Comparison the data of theoretical and numerical calculations for G-dependence of the averaged $m_y$ at different frequencies and $r=0.5$ and $\alpha=1$. Thin lines show theoretical G-dependence of $m_y$ according to formula (\ref{eq:ep1}). Symbols show numerical results, thick lines guide the eyes.}
\label{4}
\end{figure}

Table shows the values of the averaged $m_y$ obtained by analytic and numerical calculations at different $G$. The difference in  values of the averaged $m_y$ obtained by all three used methods is rather small.
\begin{table}
        \caption{$m_y$ from the theory and the numerical results}
        \label{table:my_vs_my0}
\begin{ruledtabular}
\begin{tabular}{ccccc}
$G$        &        $\Theta_0$        &  analytic $m_y$         & numerics-1  & numerics-2\\
$15.7$        &        $1.462$        &        $0.109$        &        $0.108$ &        $0.109$\\
$31.4$        &        $0.387$        &        $0.369$        &        $0.376$ &        $0.378$\\
$47.1$        &        $0.663$        &        $0.577$        &        $0.611$ &        $0.616$\\
$62.8$        &        $0.859$        &        $0.689$        &        $0.748$ &        $0.756$\\
$157$        &        $1.272$        &        $0.817$        &        $0.945$ &        $0.956$\\
\end{tabular}
\end{ruledtabular}
\end{table}
There is an essential difference between the original Kapitza pendulum and our system.  In Kapitza pendulum the stability of new equilibrium point is determined by amplitude and frequency of external force. In our case, two new points are always stable and their positions on the sphere is determined by parameters of the system.

\begin{figure}[h!]
\centering
\includegraphics[height=45mm]{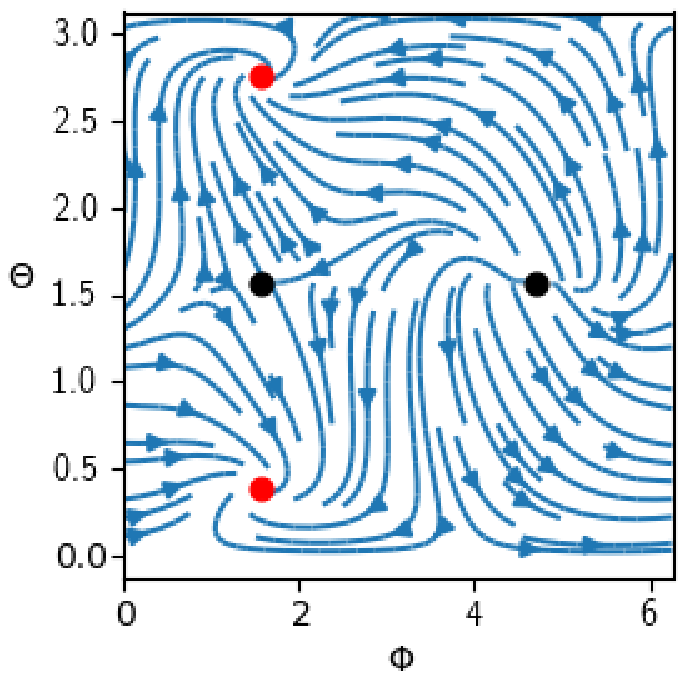}\includegraphics[height=45mm]{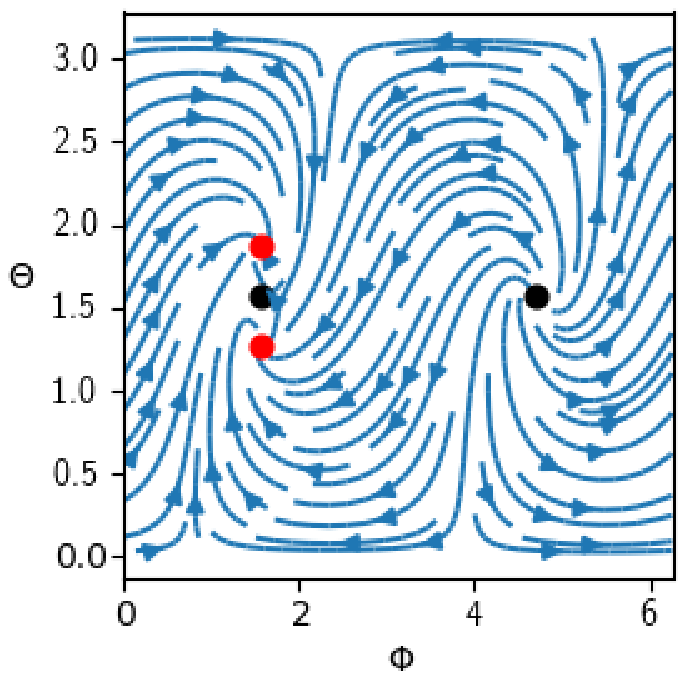}
\caption{ Phase planes (\ref{eq:main1}) at  $G= 10 \pi$ (left) and $G=50\pi$ (right). Here $r = 0.5$, $\alpha = 1$, $\omega = 70$. Red points are stable equilibrium points which are calculated from (\ref{eq:ep1}), black points are unstable (\ref{eq:ep0}). With increase in $G$ stable red points are approaching unstable ones $\Phi_0 = \pi/2$, $\Theta_0 = \pi/2$.}
\label{5}
\end{figure}

In our case, two new points are always stable and there positions on the sphere is determined by parameters of the system.  Figure \ref{5} shows vector fields in the plane $ 0 \leq \Phi \leq 2\pi$, $ 0 < \Theta < \pi$ according to the equation (\ref{eq:main1}) at two different values of $G$: $5\pi$ and  $50\pi$. It demonstrates that with an increase in $G$ two equilibrium positions are approaching the third unstable one  $\Phi = \frac{\pi}{2}$, $\Theta=0$. This result is in agreement with (\ref{eq:ep1}), which shows that an increase in  $G$ leads to the increase in $\beta$, and to the approaching of $\sin \Theta_0$ to one.

There is another interesting phenomena which is realized in our case. With changing parameters of the system, the stable equilibrium positions of  ``slow'' movement are approaching unstable one $\Phi = \frac{\pi}{2}$, $\Theta=\pi /2$. When the distance between them is getting comparable with an amplitude of ``fast'' one, all three special  points of ``slow'' system effectively are merged.

In conclusion, we have demonstrated that the coupling between magnetic moment and Josephson phase difference in $\varphi_0$ junction may effectively lead to the strong re-orientation of magnetic easy axis under the applied voltage.
This serves a manifestation of the Kapitza-like pendulum behavior and open a new way to the magnetization control by a superconducting current.

The authors thank K. Sengupta,  K. Kulikov, K. Sengupta, I. Bobkova, A. Bobkov for useful discussions. The
study was partially funded by the RFBR (research project 18-02-00318),
and the SA-JINR  collaborations. Yu. M. S. gratefully
acknowledges support from the University of South Africa's visiting
researcher program.

\begin{widetext}

\newpage

\begin{center}
\textbf{Supplement for paper Re-orientation of easy axis in $\varphi_0$ junction}

Yu. M. Shukrinov$^{1,2}$, A. Mazanik$^{1,3}$, I. R. Rahmonov$^{1,4}$, A. E. Botha$^{5}$, A. Buzdin$^{6}$
\end{center}

$^{1}$BLTP, JINR, Dubna, Moscow Region, 141980, Russia

$^{2}$Dubna State University, Dubna,  141980, Russia

$^{3}$Moscow Institute of Physics and Technology, Institutsky lane 9, Dolgoprudny, Moscow region, 141700, Russia

$^{4}$Umarov Physical Technical Institute, TAS, Dushanbe, 734063, Tajikistan

$^{5}$Department of Physics, University of South Africa, Private Bag X6, Florida 1710, South Africa

$^{6}$University Bordeaux, LOMA UMR-CNRS 5798, F-33405 Talence Cedex, France

\section{Method}

Full system of equations for magnetization:
\begin{equation}
    \begin{aligned}
        \begin{cases}
            \dot{m}_x =-m_z m_y +\Gamma m_z \sin{\left[\omega \tau-\varphi_0 \right]}+\alpha\left[ m_y \dot{m}_z - m_z \dot{m}_y \right], \\
            \dot{m}_y = m_z m_x + \alpha\left[ m_z \dot{m}_x - m_x \dot{m}_z\right], \\
            \dot{m}_z =-\Gamma m_x \sin{\left[ \omega \tau-\varphi_0\right]}+\alpha\left[m_x \dot{m}_y-m_y \dot{m}_x\right].
        \end{cases}
    \end{aligned}\label{equ-dec}
\end{equation}

The $m_{x,y,z} = M_{x,y,z}/M_0$ satisfy the constraint
$\sum_{\alpha=x,y,z} m_{\alpha}^2(t)=1$. In this system of equations
time is normalized to the inverse ferromagnetic resonance frequency
$\omega_{F}=\gamma K/M_{0}: (t\rightarrow t \omega_F)$, $\gamma$ is the gyromagnetic ratio, and $M_{0}=\|{\bf M}\|$.

System of equations of magnetic moment motion in angular variables $m_z = \cos{\theta}$, $m_x = \sin{\theta}\cos{\varphi}$, $m_y = \sin{\theta}\sin{\varphi}$ has a form:
\begin{equation}\label{eq:main}
	\begin{aligned}
		\begin{cases}
			\dot{\phi} = \frac{\cos{\theta}}{1+\alpha^2} - 	\frac{Gr}{1 + \alpha^2}\frac{1}{\sin{\theta}}\left[ \cos{\theta} \sin{\phi} - \alpha \cos{\phi}\right]\sin{\left[ \omega\tau - r \sin{\theta} \sin{\phi} \right]},\\
			\dot{\theta} =- \frac{\alpha \sin{2\theta}}{2(1+\alpha^2) } +  	\frac{Gr}{1+\alpha^2}\left[ \alpha \cos{\theta} \sin{\phi} +  \cos{\phi} \right] \sin{\left[\omega \tau - r \sin{\theta}\sin{\phi}\right]}.
		\end{cases}
	\end{aligned}
\end{equation}

We  consider that
\begin{equation} \label{eq:Method}
    \begin{aligned}
        \theta \to \Theta + \xi, \\
        \varphi \to \Phi + \eta.
    \end{aligned}
\end{equation}
Here $\Theta$ and $\Phi$ describe slow movement, while $\xi$ and  $\eta$ are coordinates for fast-varying movement. We consider $\overline{\theta} = \Theta$, $\overline{\phi} = \phi$, where averaging is taken over the period of the fast-varying force $T = \frac{2\pi}{\omega}$.

Using (\ref{eq:Method}) in the system of equations (equ. (3) in main text), and expanding on the $\xi$ and $\eta$ till members of first order, we get
\begin{equation} \label{eq:expansion}
    \begin{aligned}
       \begin{cases}
            \dot{\Phi} + \dot{\eta} \approx \frac{\cos{\Theta}}{1+\alpha^2} - \xi\frac{\sin{\Theta}}{1+\alpha^2} + F_{\phi}(\Theta, \Phi, \tau) +\\ \hspace{1 cm} \frac{ \partial F_{\phi}(\Theta, \Phi, \tau) }{\partial \Theta} \xi +  \frac{ \partial F_{\phi}(\Theta, \Phi, \tau) }{\partial \Phi} \eta, \\
            \dot{\Theta} + \dot{\xi} = \frac{\alpha \sin{2\Theta}}{2(1+\alpha^2) } + \xi \frac{\alpha \cos{2\Theta}}{(1+\alpha^2) } + F_{\theta}(\Theta, \Phi, \tau) + \\
            \hspace{1 cm} \frac{ \partial F_{\theta}(\Theta, \Phi, \tau) }{\partial \Theta} \xi +  \frac{ \partial F_{\theta}(\Theta, \Phi, \tau) }{\partial \Phi} \eta.
       \end{cases}
    \end{aligned}
\end{equation}

where

\begin{equation}
    \begin{aligned}
    \begin{cases}
        F_{\phi}(\Theta, \Phi, \tau) = - \frac{Gr}{1 + \alpha^2}\frac{1}{\sin{\Theta}}\left[ \cos{\Theta} \sin{\Phi} - \alpha \cos{\Phi}\right]\sin{\left[ \omega\tau - r \sin{\Theta} \sin{\Phi}\right]}, \\
        F_{\theta}(\Theta, \Phi, \tau) = \frac{Gr}{1+\alpha^2}\left[ \alpha \cos{\Theta} \sin{\Phi} +  \cos{\Phi} \right] \sin{\left[\omega \tau - r \sin{\Theta}\sin{\Phi}\right]}.
    \end{cases}
    \end{aligned}
\end{equation}

For the fast movement we have

\begin{equation}
    \begin{cases}
        \dot{\eta} = - \frac{Gr}{1 + \alpha^2}\frac{1}{\sin{\Theta}}\left[ \cos{\Theta} \sin{\Phi} - \alpha \cos{\Phi}\right]\sin{\left[\omega \tau - r \sin{\Theta}\sin{\Phi}\right]}, \\
        \dot{\xi} = \frac{Gr}{1+\alpha^2}\left[ \alpha \cos{\Theta} \sin{\Phi} +  \cos{\Phi} \right] \sin{\left[\omega \tau - r \sin{\Theta}\sin{\Phi}\right]},
    \end{cases}
\end{equation}

because other terms are proportional to $\xi$ or $\eta$, while $\dot{\eta} \sim \omega \eta \gg \eta$, $\dot{\xi} \sim \omega \xi \gg \xi$.

By direct integration  we see that
\begin{equation}\label{eq:fast_movement}
    \begin{aligned}
        \begin{cases}
            \eta =  \frac{Gr}{\omega}\frac{1}{1 + \alpha^2}\frac{1}{\sin{\Theta}}\left[ \cos{\Theta} \sin{\Phi} - \alpha \cos{\Phi}\right]\cos{\left[\omega \tau - r \sin{\Theta}\sin{\Phi}\right]}, \\
            \xi = -\frac{Gr}{\omega}\frac{1}{1+\alpha^2}\left[ \alpha \cos{\Theta} \sin{\Phi} +  \cos{\Phi} \right] \cos{\left[\omega \tau - r \sin{\Theta}\sin{\Phi}\right]}.
        \end{cases}
    \end{aligned}
\end{equation}
Using these $\xi$, $\eta$ in equation  \ref{eq:expansion} and do averaging over $T = \frac{2 \pi}{\omega}$, we are obtaining the system of equations for slow movement
\begin{equation} \label{eq:main1}
	\begin{aligned}
		\begin{cases}
			\dot{\Phi} = \frac{\cos{\Theta}}{1+\alpha^2}-\frac{\left(Gr\right)^2 r \alpha}{2 \omega (1+\alpha^2)^2}\cdot\frac{1}{\sin{\Theta}}[\cos{\Theta}\sin{\Phi}\\
			\hspace{1 cm}             - \alpha\cos{\Phi} ]\left\{1-\sin^2{\Theta}\sin^2{\Phi}\right\}, \\
			\dot{\Theta} =- \frac{\alpha \sin{2 \Theta}}{2(1+\alpha^2)}+\frac{\left(Gr\right)^2 r \alpha}{2 \omega (1+\alpha^2)^2}[\alpha\cos{\Theta}\sin{\Phi}\\
			\hspace{1 cm}            + \cos{\Phi} ]\left\{1-\sin^2{\Theta}\sin^2{\Phi}\right\}
		\end{cases}
	\end{aligned}
\end{equation}

\section{Stable points}

Linearization  of
(\ref{eq:main1}) after substitution $\Phi = \Phi_0 + \delta \phi,\Theta = \Theta_0 + \delta \theta$ at the equilibrium points
\begin{equation} \label{eq:ep1}
	\begin{aligned}
		\Phi_0 = \frac{ \pi}{2}, \\
		\Theta_0 = \operatorname{ArcSin}{\frac{-1 + \sqrt{1 + 4 \beta^2}}{2\beta}} \text{ or } \Theta_0 = \pi - \operatorname{ArcSin}{\frac{-1 + \sqrt{1 + 4 \beta^2}}{2\beta}}
	\end{aligned}
\end{equation}
 gives
\begin{equation} \label{eq:linmain}
    \begin{aligned}
        \begin{cases}
            \dot{\delta \phi} = - \frac{\alpha }{1 + \alpha^2}\delta \phi + \left\{ \frac{\sqrt{1+4\beta^2}}{\beta^2(1+\alpha^2)} \right\}\delta \theta, \\
            \dot{\delta \theta} =  \frac{\sqrt{1+4\beta^2}-1}{2 \beta (1+\alpha^2) }\delta\phi + \left\{ \frac{\alpha(-1-4\beta^2+\sqrt{1+4\beta^2})}{2 \beta^2 (1+\alpha^2)} \right\} \delta \theta.
        \end{cases}
    \end{aligned}
\end{equation}
Eigenvalues of this system are
\begin{equation} \label{eq:lambdas}
	\begin{aligned}
		\lambda_1 = \frac{-\sqrt{4 \alpha ^2 \beta ^4+8 \alpha ^2 \beta ^2-4 \alpha ^2 \sqrt{4 \beta ^2+1} \beta ^2-2 \alpha ^2 \sqrt{4 \beta ^2+1}+2 \alpha ^2-32 \beta ^4+8 \sqrt{4 \beta ^2+1} \beta ^2-8 \beta
		^2} }{4 \left(\alpha ^2+1\right) \beta ^2} + \\ + \frac{-6 \alpha  \beta ^2+\alpha  \sqrt{4 \beta ^2+1}-\alpha}{4 \left(\alpha ^2+1\right) \beta ^2}, \\
		\lambda_2 = \frac{\sqrt{4 \alpha ^2 \beta ^4+8 \alpha ^2 \beta ^2-4 \alpha ^2 \sqrt{4 \beta ^2+1} \beta ^2-2 \alpha ^2 \sqrt{4 \beta ^2+1}+2 \alpha ^2-32 \beta ^4+8 \sqrt{4 \beta ^2+1} \beta ^2-8 \beta
		^2} }{4 \left(\alpha ^2+1\right) \beta ^2} + \\ + \frac{-6 \alpha  \beta ^2+\alpha  \sqrt{4 \beta ^2+1}-\alpha}{4 \left(\alpha ^2+1\right) \beta ^2}.
	\end{aligned}
\end{equation}
$\lambda_1$ and $\lambda_2$ are negative. So, these points are stable. Let's test it. The denominator is always positive. $\alpha > 0$, $\beta > 0$.
\begin{equation}
	\begin{aligned}
		A = -\alpha(1 + 6 \beta^2 - \sqrt{4 \beta^2 + 1}),\\
		B = \sqrt{4 \alpha ^2 \beta ^4+8 \alpha ^2 \beta ^2-4 \alpha ^2  \beta ^2 \sqrt{4 \beta ^2+1}-2 \alpha ^2 \sqrt{4 \beta ^2+1}+2 \alpha ^2-32 \beta ^4+8  \beta ^2\sqrt{4 \beta ^2+1} -8 \beta
	^2}
	\end{aligned}
\end{equation}

We see, that imaginary part may occur only when $B$ is imaginary. If $B$ is imaginary, the points are stable because $A$ is always negative.  Let $B$ be the real, in this case we can find instability ($\lambda_{1}$ or $\lambda_{2}$ are positive) if it does exist. $A =  -\alpha(1 + 6 \beta^2 - \sqrt{4 \beta^2 + 1}) < 0$ always.  We have
\begin{equation}
	\vert A \vert^2 - \vert B \vert^2 = (\alpha^2 + 1)8\beta^2\left[1 + 4 \beta^2 - \sqrt{1 + 4\beta^2}\right] \geq 0.
\end{equation}
It means that
\begin{equation}
	\begin{aligned}
		\vert A \vert \geq \vert B \vert, \\
		-\vert A \vert + \vert B \vert \leq 0.
	\end{aligned}
\end{equation}
So, these points are stable equilibrium point.

\section{Conditions for the theory}

There are two ways to formulate conditions when our theory works. In the first way we imply that adding to angles $\Theta$ and $\Phi$, arriving from the fast movement are small in comparison with $1$. So, we need to estimate $\xi$ and $\eta$ on the trajectory of slow movement. To do it, we need $\Theta(\tau)$ and $\Phi(\tau)$. From the phase portrait (see Fig. 5 in the main text) we see that starting from an arbitrary point on the sphere,  the magnetic moment tends to line up towards to one of the stable points, so the trajectory is attracted by these points. We may evaluate $\xi$ and $\eta$ near the equilibrium points  and make them to be small in comparison to $1$.
\begin{equation}\label{eq:cond}
	\begin{aligned}
		\begin{cases}
			\eta =   \frac{Gr}{\omega}\frac{1}{1 + \alpha^2}\frac{1}{\sin{\Theta}}\left[ \cos{\Theta} \sin{\Phi} - \alpha \cos{\Phi}\right]\cos{\left[\omega \tau - r \sin{\Theta}\sin{\Phi}\right]} \bigg\vert_{\Theta = \Theta_0\text{, } \Phi = \Phi_0}  \sim  \frac{Gr}{\omega}\frac{1}{1 + \alpha^2}\frac{\cos{\Theta_0}}{\sin{\Theta_0}} \ll 1, \\
			\xi = -\frac{Gr}{\omega}\frac{1}{1+\alpha^2}\left[ \alpha \cos{\Theta} \sin{\Phi} +  \cos{\Phi} \right] \cos{\left[\omega \tau - r \sin{\Theta}\sin{\Phi}\right]} \bigg\vert_{\Theta = \Theta_0\text{, } \Phi = \Phi_0} \sim \frac{Gr}{\omega}\frac{\alpha \cos{\Theta_0}}{1+\alpha^2} \ll 1.
		\end{cases}
	\end{aligned}
\end{equation}

The second approach is to calculate the fast movement changes $m_x(\tau)$, $m_y(\tau)$, $m_z(\tau)$ near equilibrium points and make terms arriving from the fast movement slightly deflect the moment from the equilibrium points. Let's consider $m_y(\tau)$ near stable $m_{y0}$, $\Phi_0 = \pi/2$, $\sin{\Theta_0} = \frac{-1+\sqrt{1 + 4 \beta^2}}{2\beta}$, $\xi$ and $\eta$ written in (\ref{eq:fast_movement}). So, we have
\begin{equation}
\begin{aligned}
m_{y} = \sin{(\Theta_0 + \xi)}\sin{(\Phi_0 + \eta)} = \\ =\left[\sin{\Theta_0} \cos{\xi} + \cos{\Theta_0} \sin{\xi} \right] \left[\sin{\Phi_0} \cos{\eta} + \cos{\Phi_0} \sin{\eta}\right]= \\ =\sin{\Theta_0} \cos{\xi}\cos{\eta} +  \cos{\Theta_0}\sin{\xi}\cos{\eta}.
\end{aligned}
\end{equation}
We are interested in the situation when the fast movement slightly changes the slow movement. It means that we should imply $\overline{\cos{\xi}\cos{\eta}} \approx 1$ and $\overline{\sin{\xi}\cos{\eta}} \approx 0$ when the averaging is taken over the period $2\pi/\omega$. It may be read as
\begin{equation}
\begin{aligned}
\overline{\cos{\xi}\cos{\eta}} = \frac{1}{2\pi/\omega}\int_{0}^{2\pi/\omega} d\tau \cos{\left[-\frac{Gr}{\omega}\frac{\alpha \cos{\Theta_0}}{1+\alpha^2}\cos{\left[\omega \tau - r \sin{\Theta_0}\right]}\right]}\times \\ \times \cos{\left[\frac{Gr}{\omega}\frac{1}{1 + \alpha^2}\frac{\cos{\Theta_0}}{\sin{\Theta_0}}  \cos{\left[\omega \tau - r \sin{\theta}\sin{\phi}\right]}\right]} = \\
= \frac{1}{2\pi}\int_{0}^{2 \pi} dy \cos{\left[\frac{Gr}{\omega}\frac{\alpha \cos{\Theta_0}}{1+\alpha^2}\cos{y}\right]} \cos{\left[\frac{Gr}{\omega}\frac{1}{1 + \alpha^2}\frac{\cos{\Theta_0}}{\sin{\Theta_0}}  \cos{y}\right]} = \\
 = \frac{1}{2\pi}\int_{0}^{2 \pi} dy \cos{\left[\mathfrak{A}\cos{y}\right]} \cos{\left[\mathfrak{B}  \cos{y}\right]} =\frac{1}{2}\left[J_0(\mathfrak{A} - \mathfrak{B}) + J_0(\mathfrak{A}+ \mathfrak{B})\right] = \\
= \frac{1}{2} \left[ J_0\left(\frac{Gr}{\omega}\frac{\alpha \cos{\Theta_0}}{1+\alpha^2} - \frac{Gr}{\omega}\frac{1}{1 + \alpha^2}\frac{\cos{\Theta_0}}{\sin{\Theta_0}}  \right) +  J_0\left(\frac{Gr}{\omega}\frac{\alpha \cos{\Theta_0}}{1+\alpha^2} + \frac{Gr}{\omega}\frac{1}{1 + \alpha^2}\frac{\cos{\Theta_0}}{\sin{\Theta_0}} \right)\right] \approx 1.
\end{aligned}
\end{equation}
Here $\mathfrak{A} = \frac{Gr}{\omega}\frac{\alpha \cos{\Theta_0}}{1+\alpha^2}$,  $\mathfrak{B} =\frac{Gr}{\omega}\frac{1}{1 + \alpha^2}\frac{\cos{\Theta_0}}{\sin{\Theta_0}}$.
In the same way one may notice that $\overline{\cos{\xi}\sin{\eta}} = \overline{\sin{\xi}\cos{\eta}} = 0$ ($\xi(\tau)$ and $\eta(\tau)$ are $2\pi/\omega$-periodical functions in $\tau$).

Also, for $z$ direction we have to calculate new condition
\begin{equation}
\begin{aligned}
m_z = \cos{(\Theta_0 + \xi)} = \cos{\Theta_0}\cos{\xi} + \sin{\Theta_0} \sin{\xi}
\end{aligned}
\end{equation}
So, we have to imply $\overline{\cos{\xi}} \approx 1$, $\overline{\sin{\xi}} \approx 0$.
\begin{equation}
\begin{aligned}
\overline{\cos{\xi}} = \frac{1}{2\pi/\omega}\int_{0}^{2\pi/\omega}d\tau \cos{\left[-\frac{Gr}{\omega}\frac{\alpha \cos{\Theta_0}}{1+\alpha^2}\cos{\left[\omega \tau - r \sin{\Theta_0}\right]}\right]} =  \\
= J_0\left(\frac{Gr}{\omega}\frac{\alpha \cos{\Theta_0}}{1+\alpha^2} \right) \approx 1.
\end{aligned}
\end{equation}
\begin{equation}
\overline{\sin{\xi}} = \frac{1}{2\pi}\int_{0}^{2 \pi} dy \sin{\left(\mathfrak{A}\cos{y}\right)} = 0.
\end{equation}

For $x$ direction
\begin{equation}
\begin{aligned}
m_x = \sin{\left(\Theta_0 + \xi\right)\cos{\left(\Phi_0 + \eta\right)}} = \\ = \sin{\Theta_0}\cos{\xi}\sin{\eta} + \cos{\Theta_0}\sin{\xi}\sin{\eta}.
\end{aligned}
\end{equation}
So,
\begin{equation}
\begin{aligned}
\left| \overline{ \sin{\xi}\sin{\eta}} \right| = \\
= \frac{1}{2} \left| J_0\left(\frac{Gr}{\omega}\frac{\alpha \cos{\Theta_0}}{1+\alpha^2} - \frac{Gr}{\omega}\frac{1}{1 + \alpha^2}\frac{\cos{\Theta_0}}{\sin{\Theta_0}}  \right) - J_0\left(\frac{Gr}{\omega}\frac{\alpha \cos{\Theta_0}}{1+\alpha^2} + \frac{Gr}{\omega}\frac{1}{1 + \alpha^2}\frac{\cos{\Theta_0}}{\sin{\Theta_0}} \right) \right| \ll 1.
\end{aligned}
\end{equation}
%\mathfrak{A} = \frac{Gr}{\omega}\frac{\alpha \cos{\Theta_0}}{1+\alpha^2}, \text{ } \mathfrak{B} =\frac{Gr}{\omega}\frac{1}{1 + \alpha^2}\frac{\cos{\Theta_0}}{\sin{\Theta_0}},\\
So, new conditions are
\begin{equation}\label{eq:cond1}
\begin{aligned}
\begin{cases}
\frac{1}{2}\left[J_0(\mathfrak{A} - \mathfrak{B}) + J_0(\mathfrak{A}+ \mathfrak{B})\right] \approx 1, \\
J_0\left(\mathfrak{A}\right) \approx 1,\\
\frac{1}{2}\left| J_0(\mathfrak{A} - \mathfrak{B}) - J_0(\mathfrak{A}+ \mathfrak{B}) \right|  \ll 1.
\end{cases}
\end{aligned}
\end{equation}

One should notice that conditions (\ref{eq:cond1}) are fulfilled, when $\mathfrak{A}$ and $\mathfrak{B}$ are small. It implies
\begin{equation}
\begin{cases}
\begin{aligned}
\mathfrak{A} = \frac{Gr}{\omega}\frac{\alpha \cos{\Theta_0}}{1+\alpha^2} \ll 1, \\
\mathfrak{B} =\frac{Gr}{\omega}\frac{1}{1 + \alpha^2}\frac{\cos{\Theta_0}}{\sin{\Theta_0}} \ll 1.
\end{aligned}
\end{cases}
\end{equation}

We see that these conditions are coincide with (\ref{eq:cond}).

\end{widetext}

\end{document}